\documentclass[epj]{webofc}
\usepackage[varg]{txfonts}   
\woctitle{XLV International Symposium on Multiparticle Dynamics}

\newcommand{\MiNLO}{MiN\protect\scalebox{0.8}{LO}\xspace}

\newcommand{\MCatNLO}{M\protect\scalebox{0.8}{C}@N\protect\scalebox{0.8}{LO}\xspace}

\newcommand{\POWHEG}{P\protect\scalebox{0.8}{OWHEG}\xspace}
\newcommand{\PowhegBox}{P\protect\scalebox{0.8}{OWHEG}B\protect\scalebox{0.8}{OX}\xspace}

\newcommand{\NLOPS}{N\protect\scalebox{0.8}{LO}P\protect\scalebox{0.8}{S}\xspace}
\newcommand{\NNLOPS}{N\protect\scalebox{0.8}{NLO}P\protect\scalebox{0.8}{S}\xspace}

\newcommand{\MEPSatNLO}{M\protect\scalebox{0.8}{E}P\protect\scalebox{0.8}{S}@N\protect\scalebox{0.8}{LO}\xspace}
\newcommand{\CKKW}{CKKW\xspace}
\newcommand{\UNLOPS}{UN\protect\scalebox{0.8}{LO}P\protect\scalebox{0.8}{S}\xspace}
\newcommand{\UNNLOPS}{UN$^2$\protect\scalebox{0.8}{LO}P\protect\scalebox{0.8}{S}\xspace}
\newcommand{\FxFx}{F\protect\scalebox{0.8}{X}F\protect\scalebox{0.8}{X}\xspace}

\newcommand{\Herwigpp}{H\protect\scalebox{0.8}{ERWIG++}\xspace}

\newcommand{\Alpgen}{A\protect\scalebox{0.8}{LPGEN}\xspace}

\newcommand{\MadgraphaMCatNLO}{M\protect\scalebox{0.8}{AD}G\protect\scalebox{0.8}{RAPH}5\_\protect\scalebox{0.8}{A}M\protect\scalebox{0.8}{C}@N\protect\scalebox{0.8}{LO}\xspace}

\newcommand{\PythiaEight}{P\protect\scalebox{0.8}{YTHIA}8\xspace}

\newcommand{\Sherpa}{S\protect\scalebox{0.8}{HERPA}\xspace}

\newcommand{\Dire}{D\protect\scalebox{0.8}{IRE}\xspace}

\newcommand{\LEP}{LEP\xspace}

\newcommand{\Aleph}{ALEPH\xspace}

\newcommand{\ATLAS}{ATLAS\xspace}

\begin{document}
\title{Recent developments in Monte-Carlo Event Generators}

\author{Marek Sch\"onherr\inst{1}\fnsep\thanks{\email{marek.schoenherr@physik.uzh.ch}}}

\institute{Physik--Institut, Universit{\"a}t Z{\"u}rich, Winterthurerstrasse 190,
           CH--8057 Z{\"u}rich, Switzerland}

\abstract{%
  With Run II of the LHC having started, the need for high precision 
  theory predictions whose uncertainty matches that of the data to be 
  taken necessitated a range of new developments in Monte-Carlo Event 
  Generators. This talk will give an overview of the progress in 
  recent years in the field and what can and cannot be expected from 
  these newly written tools.
}
\maketitle

\begin{picture}(0,0)
  \put(332,195){ZU--TH 45/15}
\end{picture}

\section{Introduction}
\label{sec:intro}

\begin{figure}[t!]
  \centering
  \begin{minipage}{0.47\textwidth}
    \includegraphics[width=0.83\textwidth]{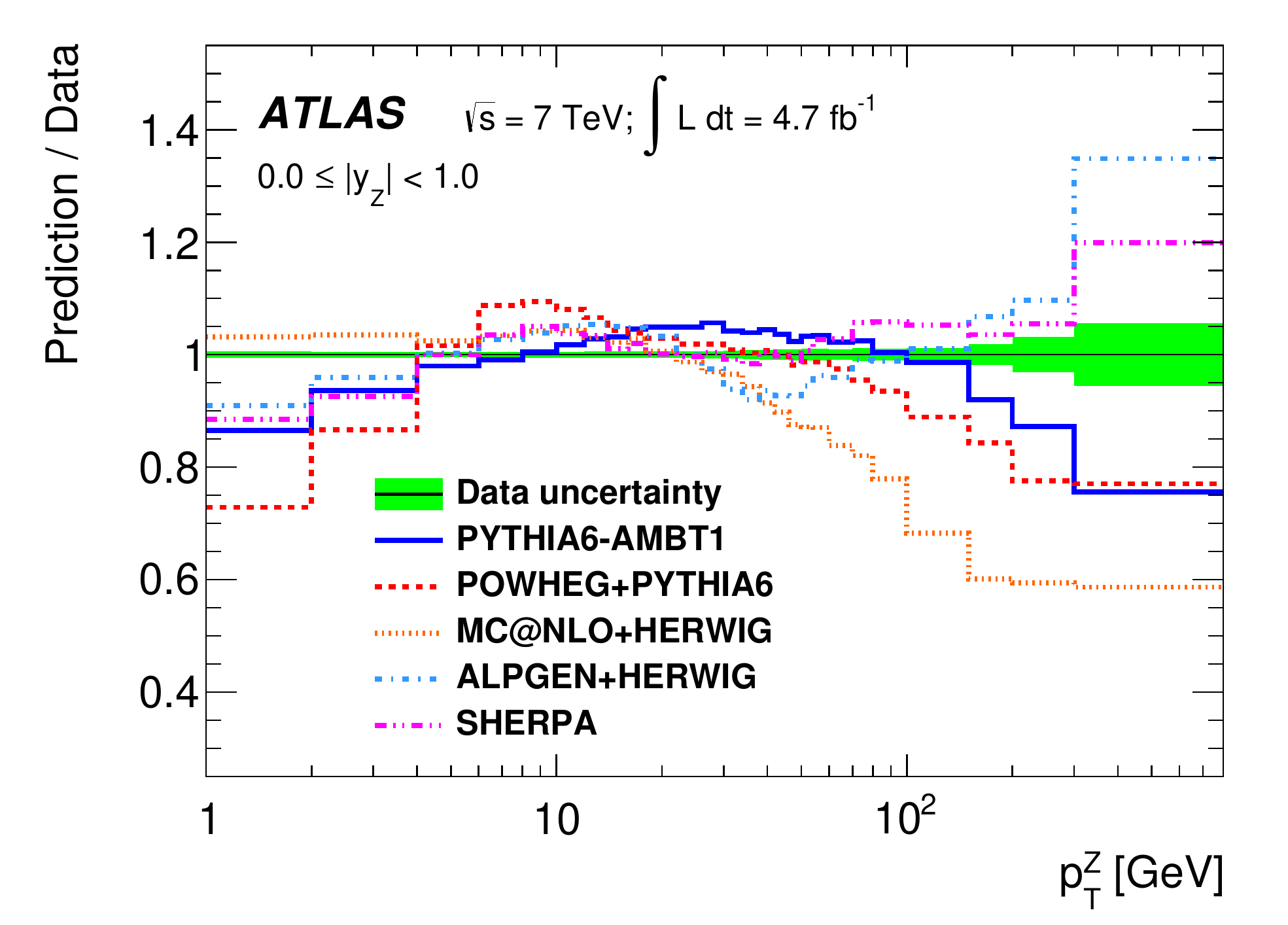}
    \includegraphics[width=0.83\textwidth]{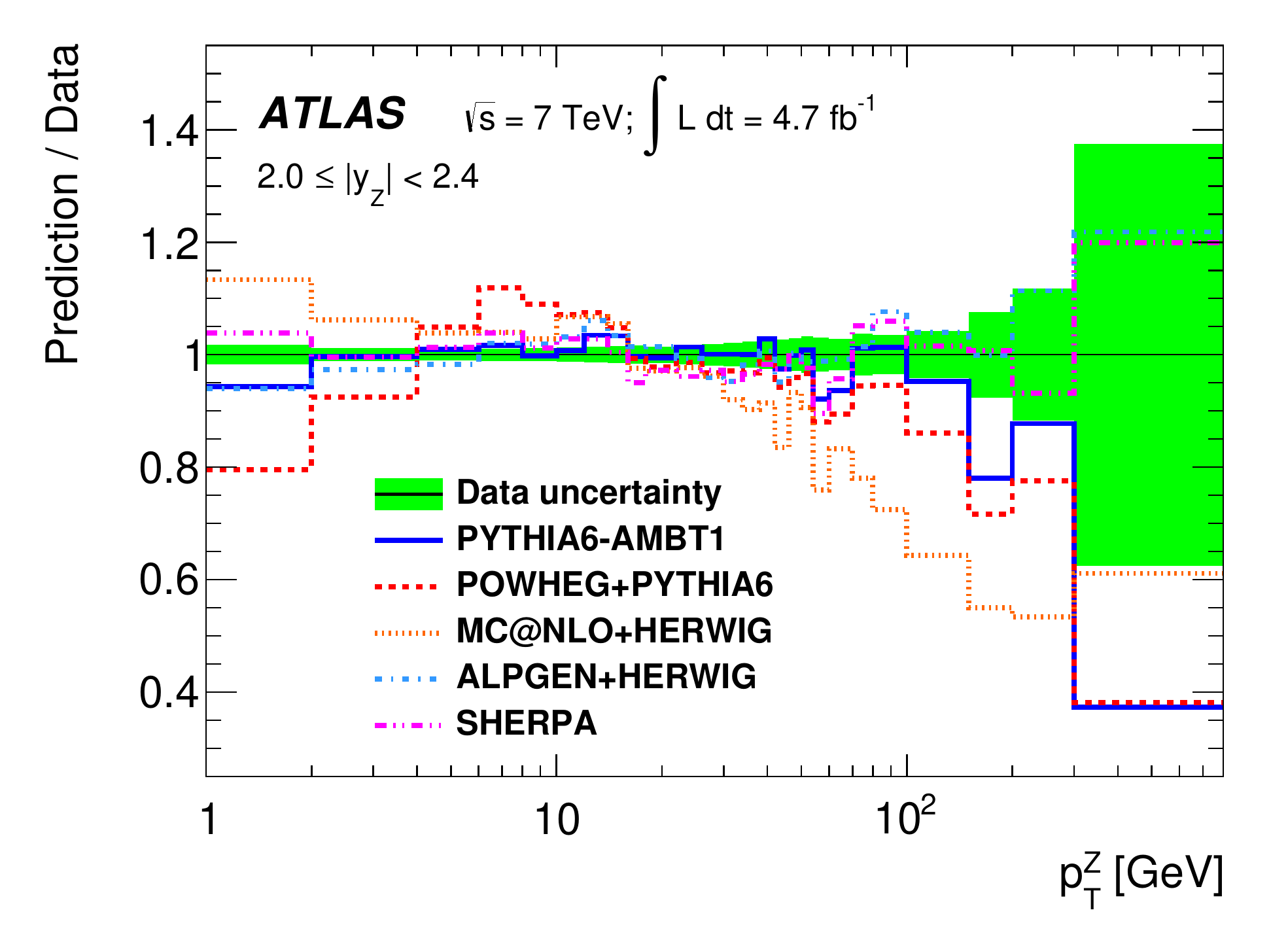}
  \end{minipage}
  \hspace*{2mm}
  \begin{minipage}{0.423\textwidth}
    \includegraphics[width=\textwidth]{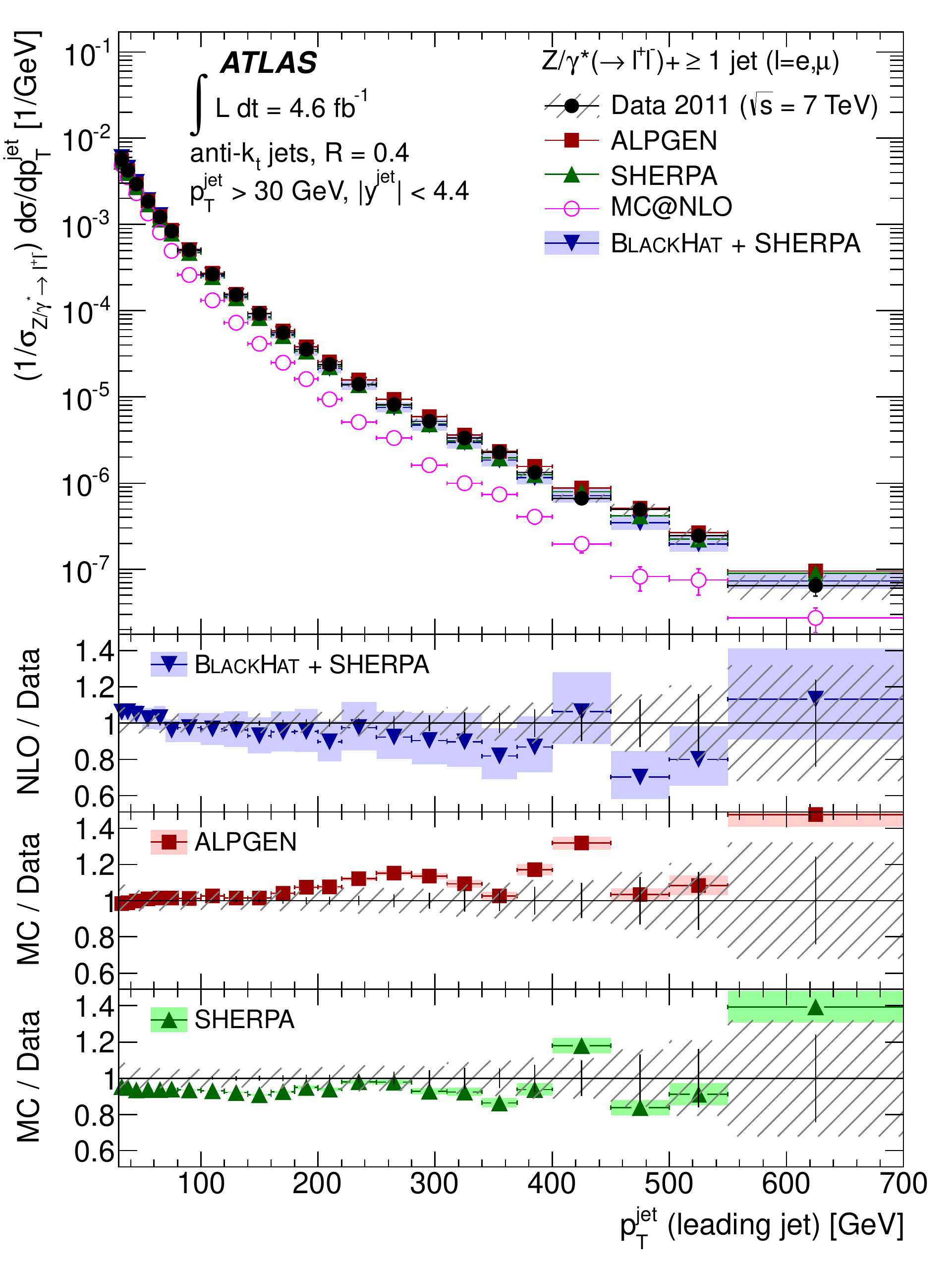}
  \end{minipage}
  \caption{
	    Left: Transverse momentum of the reconstructed $Z$ boson 
	    in the central and the forward region, as measured by the 
	    \ATLAS detector. Figure taken from \cite{Aad:2014xaa}. 
	    Right: Transverse momentum of the leading jet in $Z$ boson 
	    production in association with jets, as measured by the 
	    \ATLAS detector. Figure taken from \cite{Aad:2013ysa}.
	  }
  \label{fig:zptjpt}
\end{figure}

Modern Monte-Carlo Event Generators like \PythiaEight 
\cite{Sjostrand:2014zea}, \Herwigpp \cite{Bellm:2013hwb,Bellm:2015jjp}
and \Sherpa \cite{Gleisberg:2008ta} are instrumental in most physics 
analyses and measurements at the LHC. The current state-of-the art 
in usage at the experiments are either next-to-leading order to parton 
shower matched calculations (\NLOPS) or multijet merged ones at leading 
order accuracy. Examples for their widespread use are shown in Fig.\ 
\ref{fig:zptjpt}. In many instances the \PythiaEight and \Herwigpp 
generators (or their older predecessors) receive input from parton level 
tools computing the hard core production matrix elements either at 
NLO  for processes with few final state particles (e.g.\ \MadgraphaMCatNLO 
\cite{Alwall:2014hca} or \PowhegBox \cite{Alioli:2010xd}), or at LO for 
multileg processes (e.g.\ \Alpgen \cite{Mangano:2002ea} or \MadgraphaMCatNLO). 
The following contribution highlights a few important improvements 
thereupon effected in recent years.

\section{Parton shower developments}
\label{sec:ps}

The first avenue improvements in event generators have been 
accomplished in recent years are parton showers. Being instrumental 
for the description of many relevant observables parton showers 
are a main ingredient of all event generator frameworks and thus 
their continuing advancement is crucial to a better description 
of collider observables.

\begin{figure}[t!]
  \centering
  \includegraphics[width=0.47\textwidth]{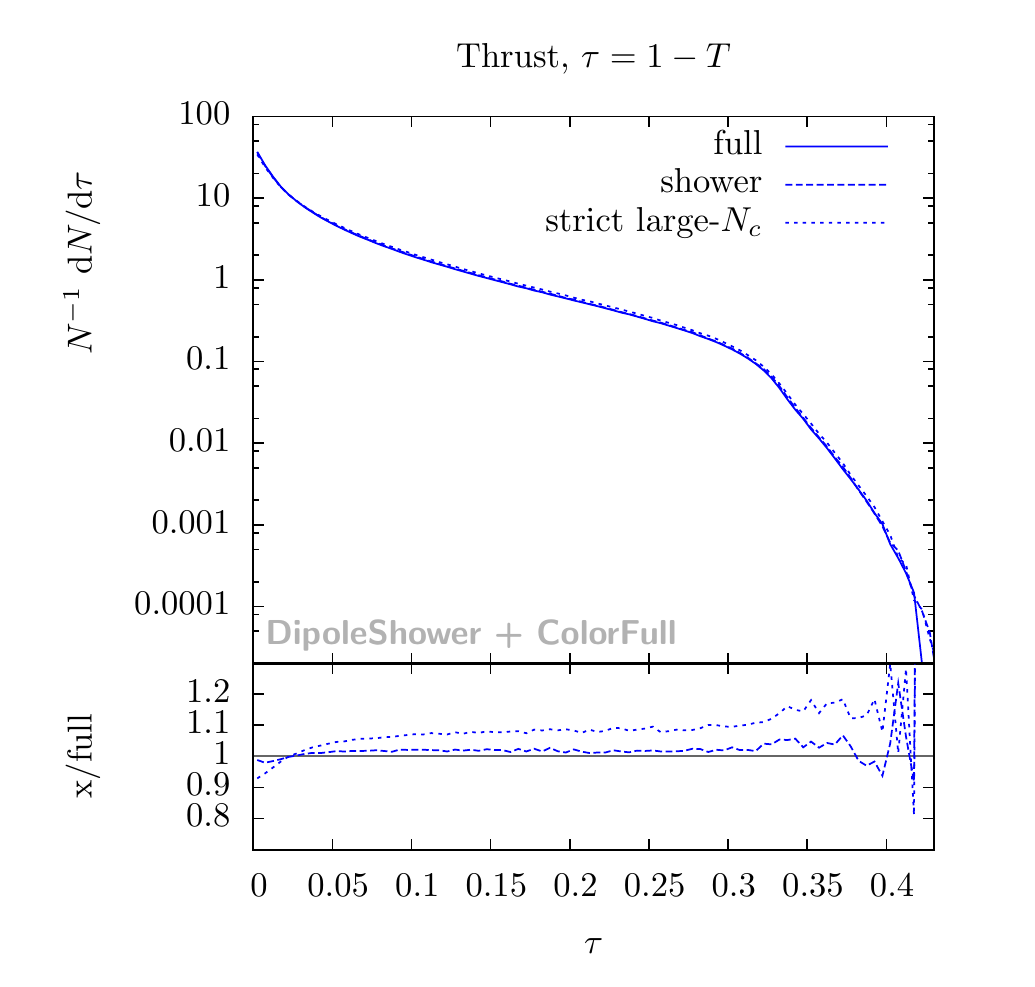}
  \hfill
  \includegraphics[width=0.47\textwidth]{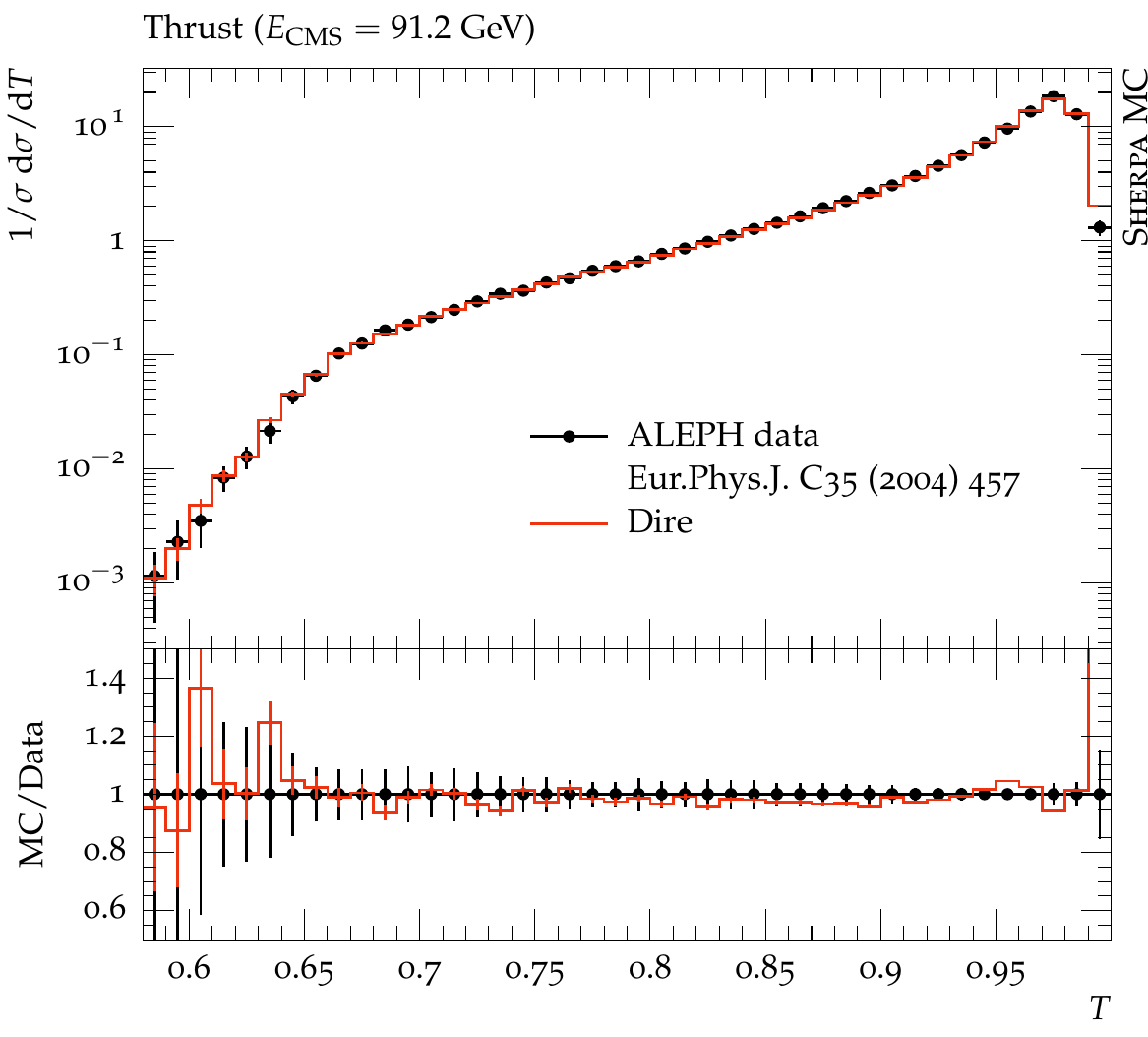}
  \caption{
            Left: Subleading colour effects in parton shower evolution in 
            thrust in $e^+e^-$-collisions at \LEP. Figure taken from 
            \cite{Platzer:2012np}. 
	    Right: Thrust in $e^+e^-$-collisions at \LEP as calculated by 
	    a new dipole shower implementation \Dire. Figure taken from 
	    \cite{Hoche:2015sya}.
	  }
  \label{fig:ps-col}
\end{figure}

On the one hand side subleading colour information has been 
propagated into the algorithms otherwise operating in the leading 
colour limit. In the first such advancement it was a pure necessity 
to achieve a process independent NLO matching and was consequently 
only introduced in the first emission \cite{Hoeche:2011fd}. Later 
implementations trace subleading colour information in different 
limits through multiple, if not all, emissions of the parton shower 
evolution \cite{Platzer:2012np,Nagy:2014mqa}. Generally, the impact 
of such improvements is small, as shown in Fig.\ \ref{fig:ps-col} (left), 
although also highly sensitive observables exist \cite{Hoeche:2013mua}.

Other works build around gaining a higher degree of analytical control 
over the parton showers' resummation properties \cite{Hoche:2015sya}. 
Through the accompanying scrutiny also their predictive power and 
ability to describe data has been improved. Fig.\ \ref{fig:ps-col} 
(right) details the results of the newly written \Dire parton shower 
as compared to \Aleph data.

\begin{figure}[t!]
  \centering
  \includegraphics[width=0.42\textwidth]{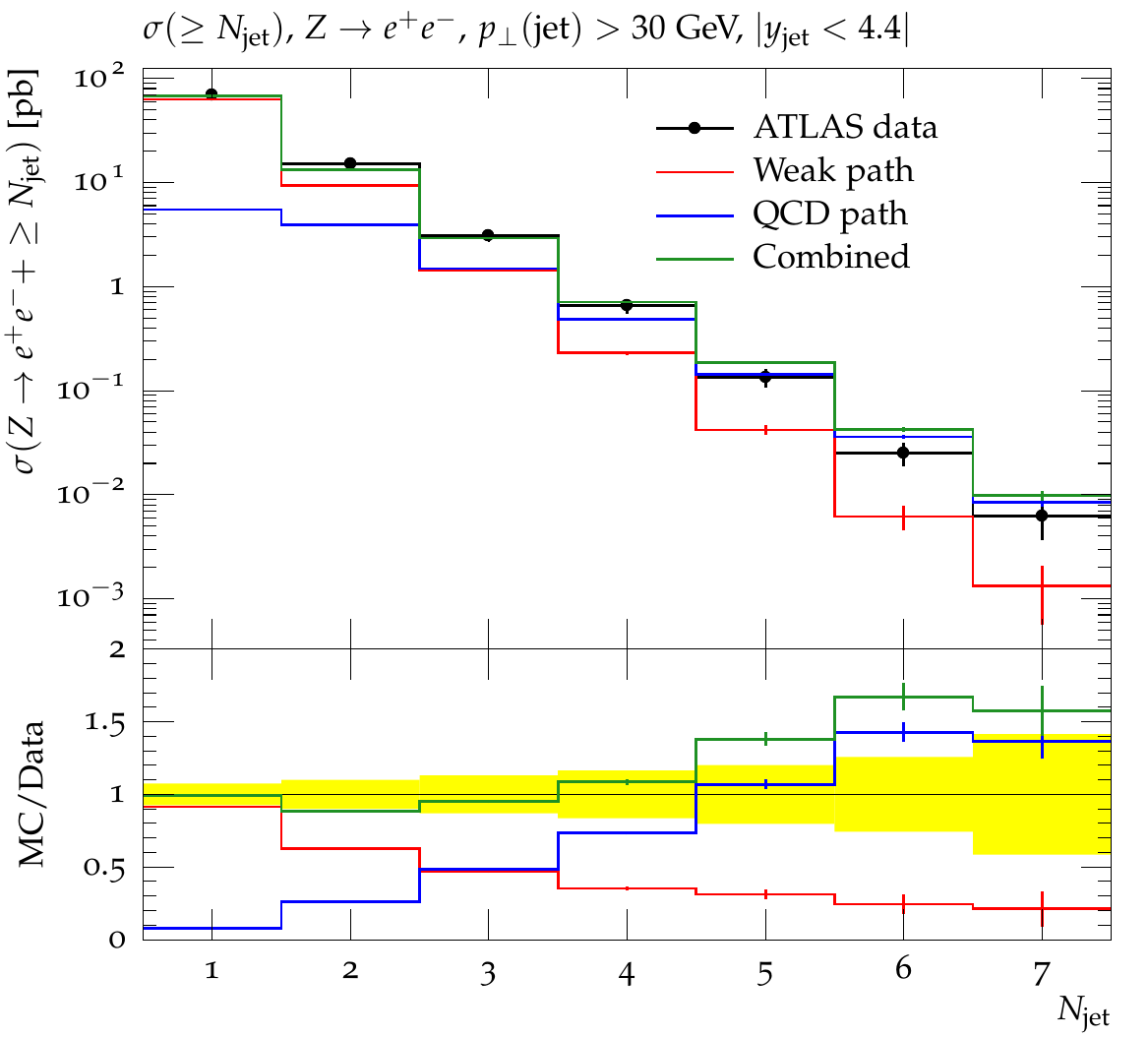}
  \hfill
  \includegraphics[width=0.5\textwidth]{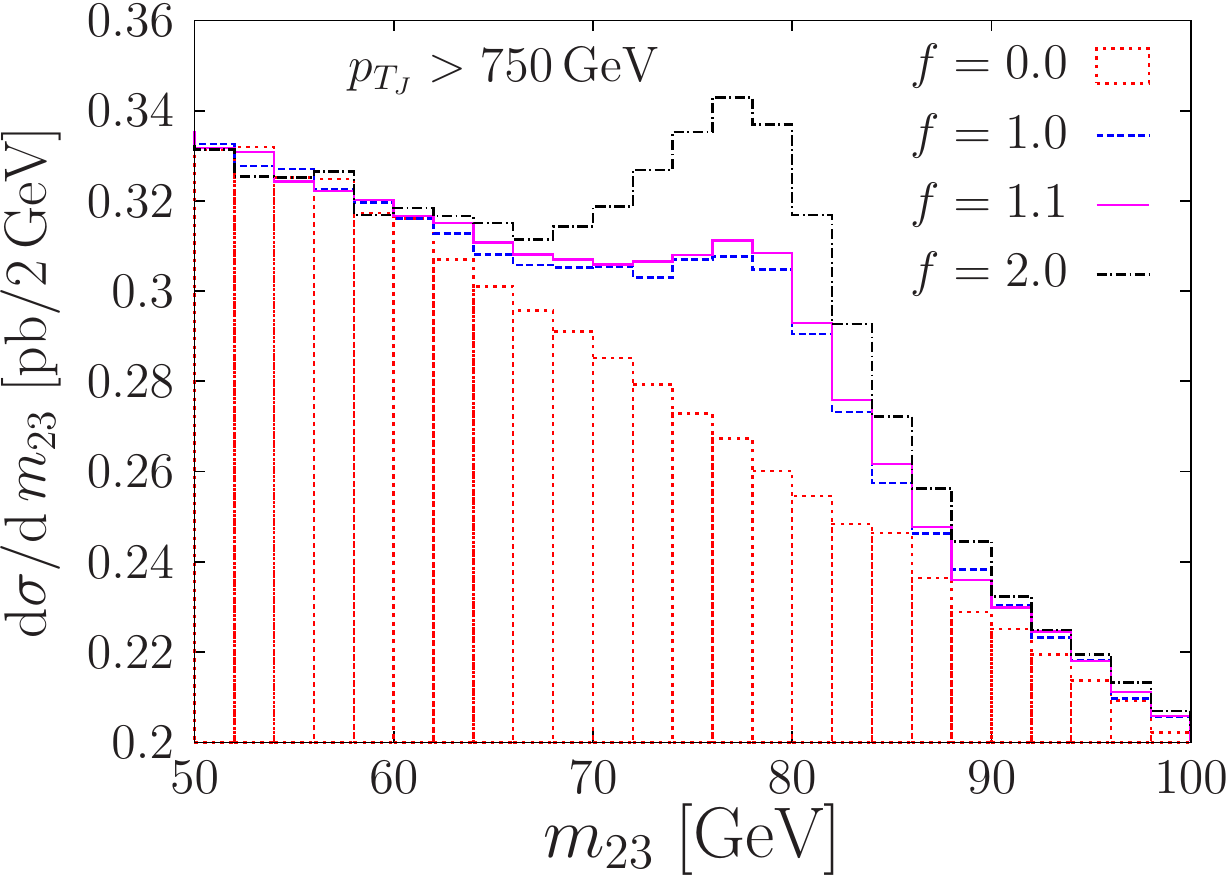}
  \caption{
            Left: Interplay of QCD evolution on top of $W$ production and 
            EW evolution on top of jet production in describing $W$ plus
            mulitjet production. Figure taken from \cite{Christiansen:2014kba}. 
	    Right: Effects of adding EW evolution on subjet invariant masses. 
	    Figure taken from \cite{Krauss:2014yaa}.
	  }
  \label{fig:ps-ew}
\end{figure}

The third stream of development centres around incorporating electroweak 
effects into parton showers \cite{Christiansen:2014kba,Krauss:2014yaa,
  Christiansen:2015jpa}. The emission of $W$ and $Z$ bosons, although 
rare, can be an important ingredient, especially in the highly boosted 
regime. Fig.\ \ref{fig:ps-ew} such effects for various observables. Such 
soft-collinear approximations to higher-order electroweak corrections 
complement the approximate NLO electroweak corrections of \cite{Gieseke:2014gka}
and the recently achieved automation of NLO electroweak corrections 
\cite{Kallweit:2014xda,Frixione:2015zaa,Kallweit:2015dum}.

\section{\protect\NLOPS matching}
\label{sec:nlops}

Known under the names of \MCatNLO \cite{Frixione:2002ik} and \POWHEG 
\cite{Nason:2004rx,Frixione:2007nw}, methods for matching NLO 
computations to parton showers are around for over ten years now. 
Recent years have seen small theoretical improvements on both schemes 
that lead to their application to a wider range of processes 
\cite{Hoeche:2011fd,Hoeche:2012ft,Hoche:2012wh,Cascioli:2013era} 
with a more complicated internal structure. The range of showers 
the respective matching schemes are available for has increased 
likewise \cite{Bellm:2013hwb,Bellm:2015jjp,Hoche:2010pf,Czakon:2015cla}. 
An systematically different matching method, \UNLOPS, was developed in 
\cite{Lonnblad:2012ix}.

\begin{figure}[t!]
  \centering
  \includegraphics[width=0.47\textwidth]{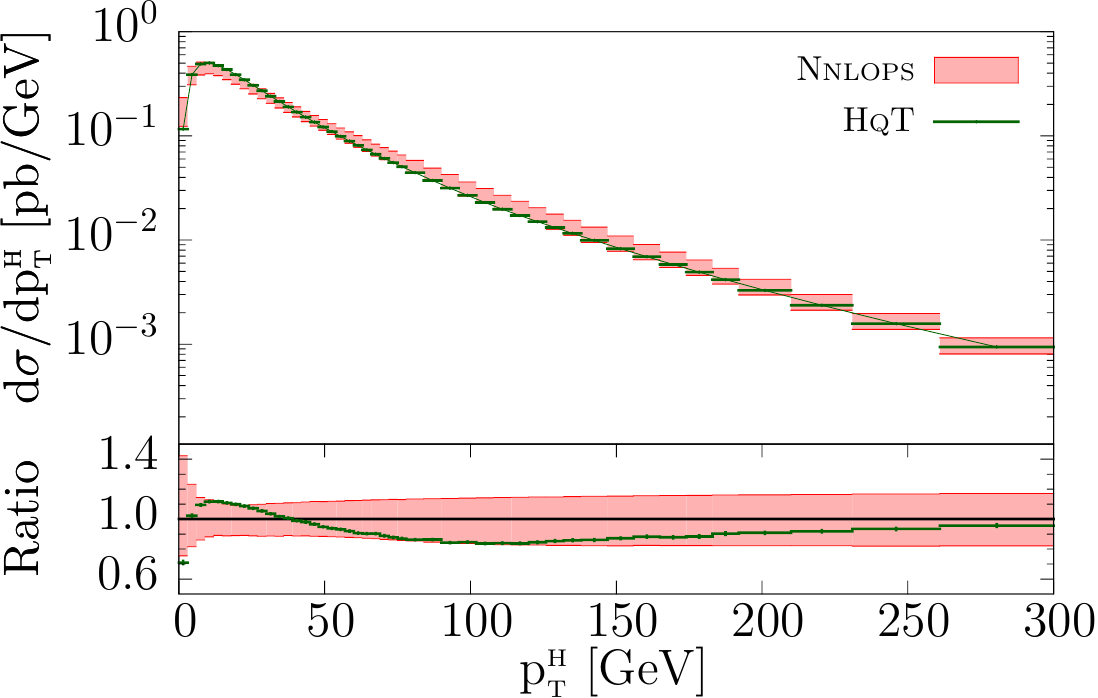}
  \hfill
  \includegraphics[width=0.47\textwidth]{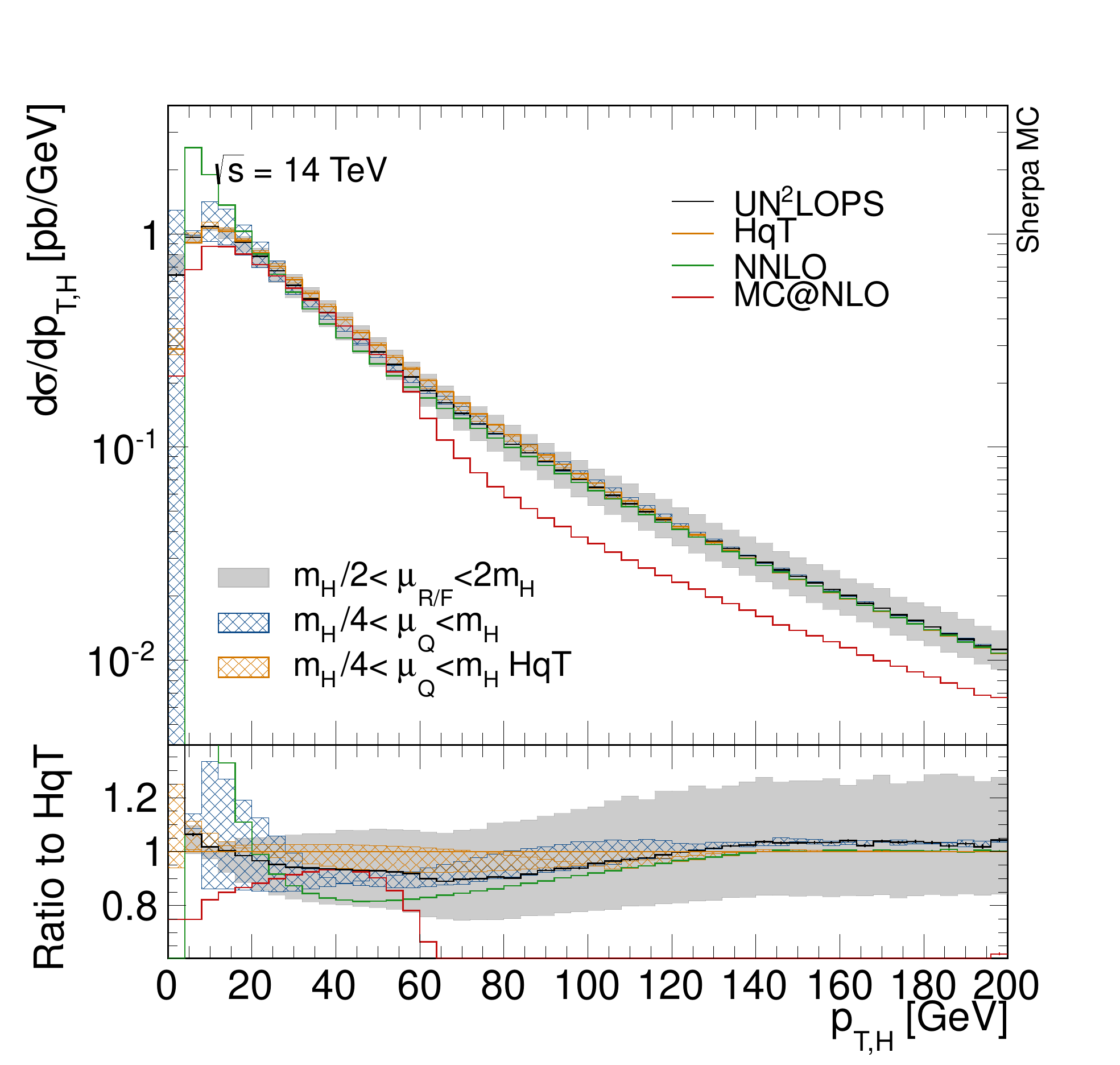}
  \caption{
            Left: Transverse momentum of the Higgs boson described at 
            \NNLOPS in the \MiNLO approach. Figure taken from 
            \cite{Hamilton:2013fea}. 
	    Right: Transverse momentum of the Higgs boson described at 
            \NNLOPS in the \UNNLOPS approach. Figure taken from 
            \cite{Hoche:2014dla}.
	  }
  \label{fig:nnlops}
\end{figure}

Similarly, \CKKW \cite{Catani:2001cc} method of scale setting and Sudakov 
factor inclusion has been elevated to be applicable to NLO QCD computations 
in \cite{Hamilton:2012np}, leading to an improvement of \NLOPS matched 
computations incorporating jets in the final state already at Born level. 
In colour singlet production in association with one additional jet the 
inclusion of a proper process dependent finite term can restore NLO 
accuracy for inclusive singlet production as well \cite{Hamilton:2012rf}. 
This formed the basis for the development of a \NNLOPS matching method 
for colour singlet production \cite{Hamilton:2013fea,Karlberg:2014qua}. 
An exemplary result is shown in Fig.\ \ref{fig:nnlops} (left). Another 
\NNLOPS matching scheme basing basing on \MCatNLO and \UNLOPS matching 
was developed for the same process class in \cite{Hoeche:2014aia,
  Hoche:2014dla}. Fig.\ \ref{fig:nnlops} (right) details the results 
for this scheme named \UNNLOPS.

\section{Multijet merging}
\label{sec:merging}

\begin{figure}[t!]
  \centering
  \begin{minipage}{0.47\textwidth}
    \begin{picture}(0,0)
      \put(-20,-80){\includegraphics[width=1.3\textwidth]{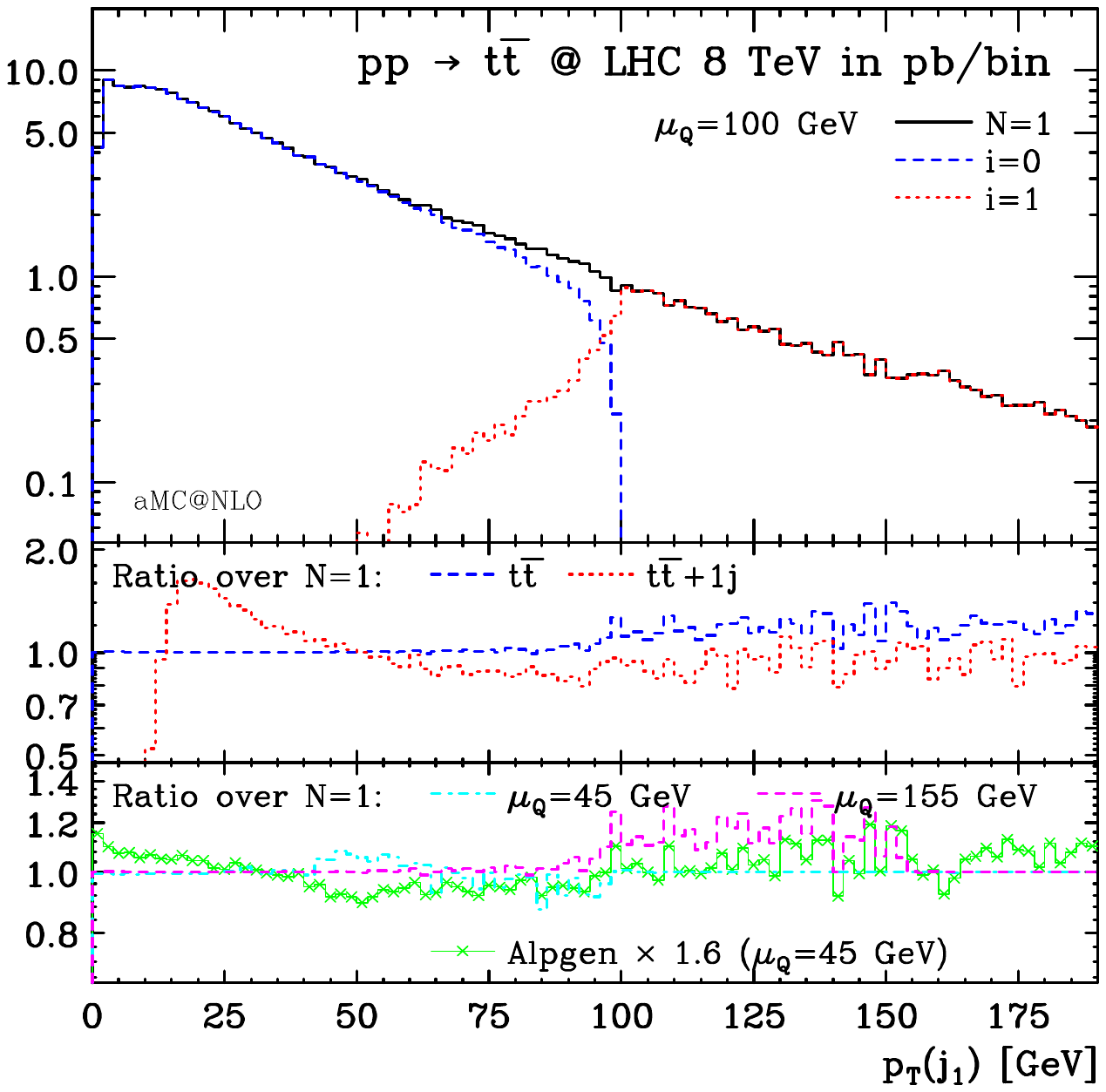}}
      \put(18,-443){\includegraphics[width=1.8\textwidth]{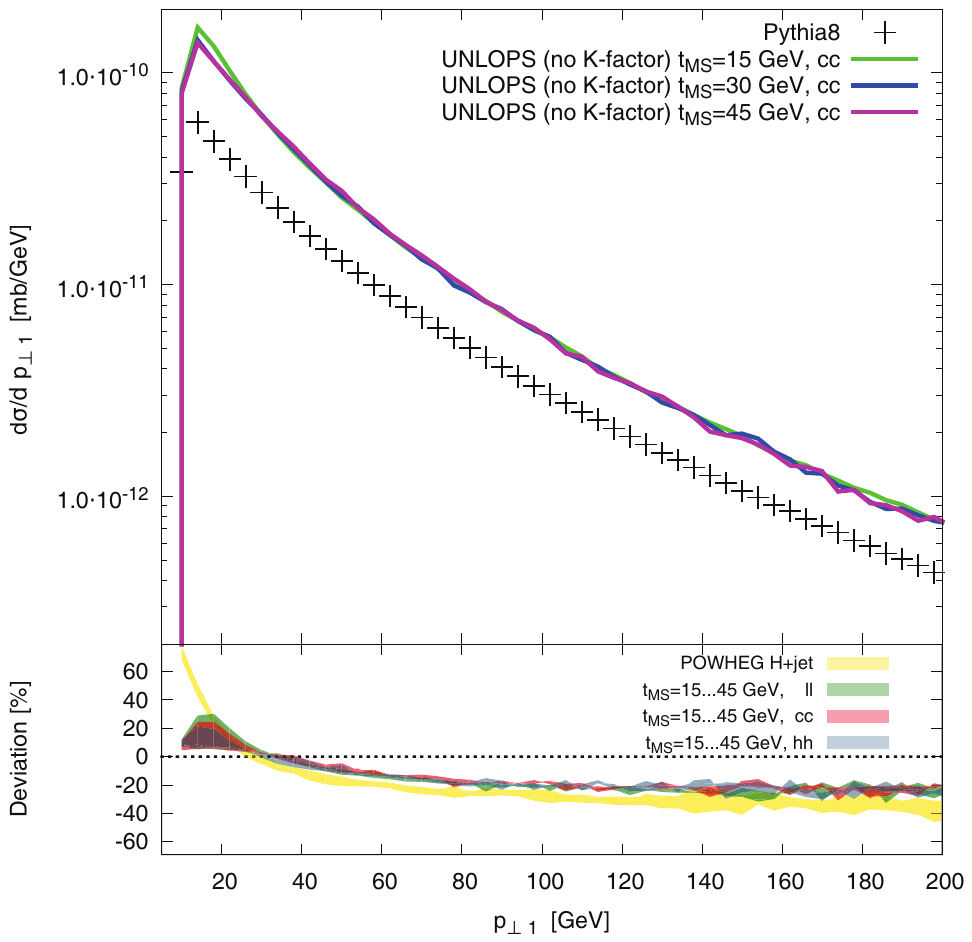}}
    \end{picture}
  \end{minipage}
  \hfill
  \begin{minipage}{0.47\textwidth}
      \lineskip-1.7pt
      \includegraphics[width=\textwidth]{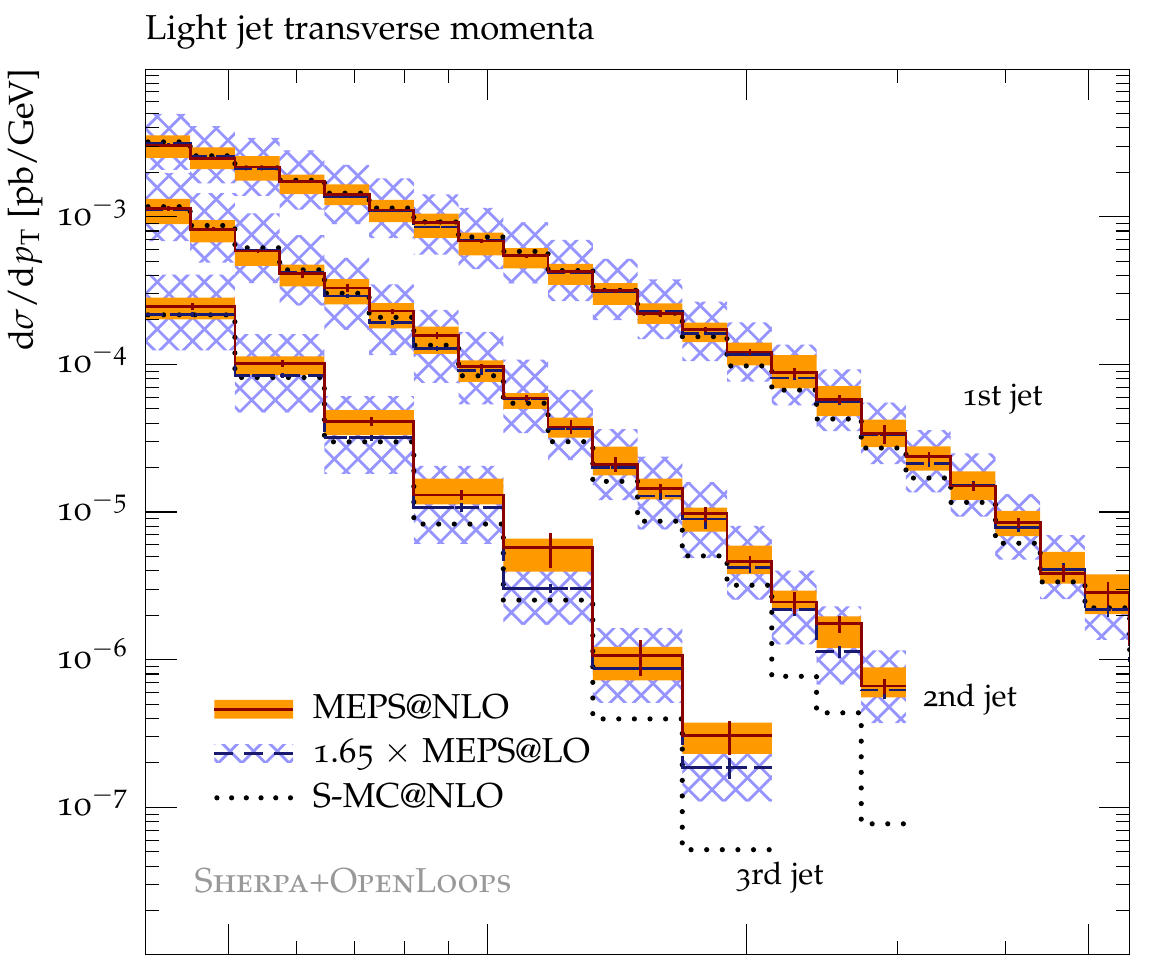}
      \includegraphics[width=\textwidth]{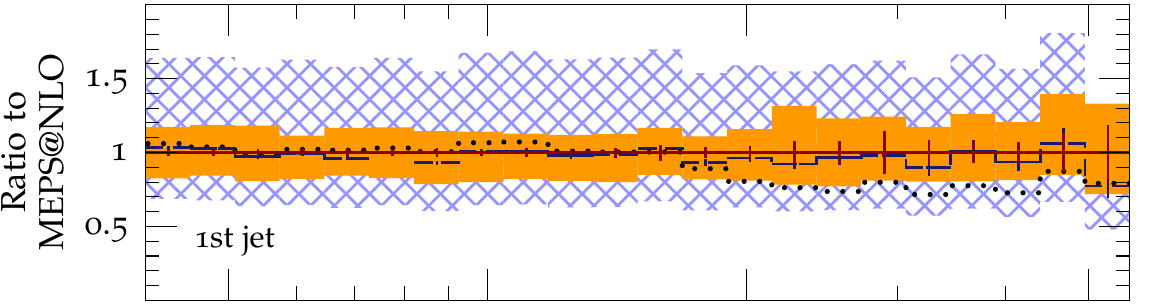}
      \includegraphics[width=\textwidth]{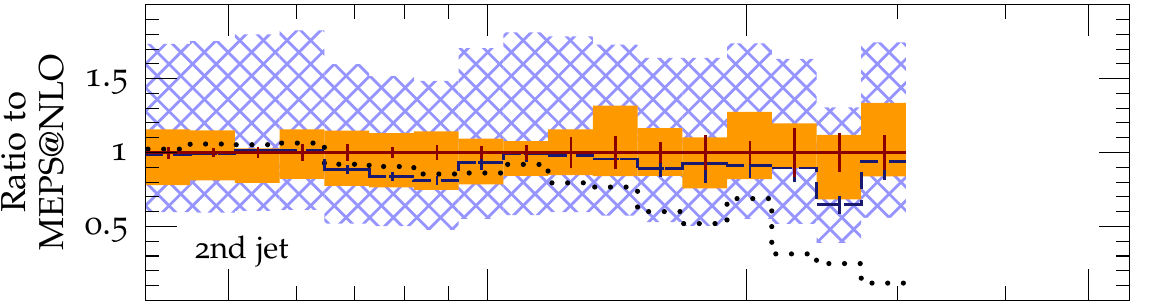}
      \includegraphics[width=\textwidth]{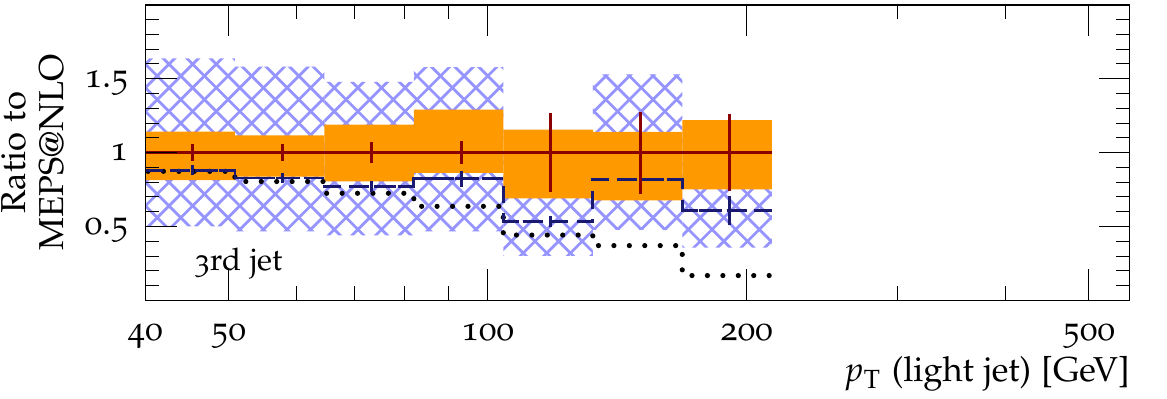}
  \end{minipage}
  \caption{
            Left top: Transverse momentum of the leading jet in top pair 
            production in association with jets described through a 
            \FxFx combination. Figure taken from \cite{Frederix:2012ps}. 
            Left bottom: Transverse momentum of the leading jet in Higgs 
            production in association with jets described through 
            \UNLOPS merging. Figure taken from \cite{Lonnblad:2012ix}. 
	    Right: Transverse momentum of the three leading jets in 
	    top pair production in association with jets described 
	    through \MEPSatNLO merging. Figure taken from 
            \cite{Hoeche:2014qda}.
	  }
  \label{fig:merge}
\end{figure}

Multijet merging aims at consistently combining calculations for the 
production of a certain experimental signature, like lepton pairs, 
Higgs bosons or top quark pairs, in association with any number of jets. 
As many observables do not clearly separate between different jet 
multiplicities but instead receive substantial contributions by e.g.\ 
one, two and three jet final states, such multijet merging schemes 
are the best way to calculate these observables with the highest 
accuracies.

At the NLO, this was pioneered in \cite{Lavesson:2008ah}. Modern 
implementations for hadron colliders first appeared as \MEPSatNLO
\cite{Hoeche:2012yf,Gehrmann:2012yg,Hoche:2010kg} and were applied 
to a wide range of processes \cite{Hoeche:2013mua,Cascioli:2013gfa,
  Hoeche:2014lxa,Hoeche:2014rya,Hoeche:2014qda,Buschmann:2014sia}. 
Other implementations using other methods to calculated the matched 
processes for each jet multiplicity have been established in 
\cite{Frederix:2012ps} and \cite{Lonnblad:2012ix}. 
Fig.\ \ref{fig:merge} details results of all three mentioned 
methods.

\section{Conclusions}
\label{sec:conclusions}

Monte-Carlo Event Generators are in good shape for Run II of the LHC. 
Tremendous progress in terms of the achieved accuracy in calculating 
the hard scattering process has been achieved. They can thus be used 
as for precise theoretical predictions including an evaluation of the 
theoretical uncertainty. Developments for the non-perturbative 
component of high-energy collisions, however, remain sparse. In that 
regime, playing a role in every hadron collider event, still 
phenomenologically motivated models with a large number of to-be-tuned 
parameters are instrumental in all generators. Thus, for precision 
calculations one should still try to minimise the influence of that 
regime on the considered observables.

MS acknowledges funding by the Swiss National Science Foundation (SNF) 
under contract PP00P2-128552.

\bibliography{refs}

\begin{thebibliography}{47}

\bibitem{Sjostrand:2014zea}
T.~Sj{\"o}strand, S.~Ask, J.R. Christiansen, R.~Corke, N.~Desai, P.~Ilten,
  S.~Mrenna, S.~Prestel, C.O. Rasmussen, P.Z. Skands, Comput. Phys. Commun.
  \textbf{191}, 159 (2015), \texttt{1410.3012}

\bibitem{Bellm:2013hwb}
J.~Bellm et~al. (2013), \texttt{1310.6877}

\bibitem{Bellm:2015jjp}
J.~Bellm et~al. (2015), \texttt{1512.01178}

\bibitem{Gleisberg:2008ta}
T.~Gleisberg, S.~H{\"o}che, F.~Krauss, M.~Sch{\"o}nherr, S.~Schumann,
  F.~Siegert, J.~Winter, JHEP \textbf{02}, 007 (2009), \texttt{0811.4622}

\bibitem{Alwall:2014hca}
J.~Alwall, R.~Frederix, S.~Frixione, V.~Hirschi, F.~Maltoni, O.~Mattelaer, H.S.
  Shao, T.~Stelzer, P.~Torrielli, M.~Zaro, JHEP \textbf{07}, 079 (2014),
  \texttt{1405.0301}

\bibitem{Alioli:2010xd}
S.~Alioli, P.~Nason, C.~Oleari, E.~Re, JHEP \textbf{06}, 043 (2010),
  \texttt{1002.2581}

\bibitem{Mangano:2002ea}
M.L. Mangano, M.~Moretti, F.~Piccinini, R.~Pittau, A.D. Polosa, JHEP
  \textbf{07}, 001 (2003), \texttt{hep-ph/0206293}

\bibitem{Aad:2014xaa}
G.~Aad et~al. (ATLAS), JHEP \textbf{09}, 145 (2014), \texttt{1406.3660}

\bibitem{Aad:2013ysa}
G.~Aad et~al. (ATLAS), JHEP \textbf{07}, 032 (2013), \texttt{1304.7098}

\bibitem{Hoeche:2011fd}
S.~H{\"o}che, F.~Krauss, M.~Sch{\"o}nherr, F.~Siegert, JHEP \textbf{09}, 049
  (2012), \texttt{1111.1220}

\bibitem{Platzer:2012np}
S.~Pl{\"a}tzer, M.~Sj{\"o}dahl, JHEP \textbf{07}, 042 (2012),
  \texttt{1201.0260}

\bibitem{Nagy:2014mqa}
Z.~Nagy, D.E. Soper, JHEP \textbf{06}, 097 (2014), \texttt{1401.6364}

\bibitem{Hoeche:2013mua}
S.~H{\"o}che, J.~Huang, G.~Luisoni, M.~Sch{\"o}nherr, J.~Winter, Phys. Rev.
  \textbf{D88}, 014040 (2013), \texttt{1306.2703}

\bibitem{Hoche:2015sya}
S.~H{\"o}che, S.~Prestel, Eur. Phys. J. \textbf{C75}, 461 (2015),
  \texttt{1506.05057}

\bibitem{Christiansen:2014kba}
J.R. Christiansen, T.~Sj{\"o}strand, JHEP \textbf{04}, 115 (2014),
  \texttt{1401.5238}

\bibitem{Krauss:2014yaa}
F.~Krauss, P.~Petrov, M.~Sch{\"o}nherr, M.~Spannowsky, Phys. Rev. \textbf{D89},
  114006 (2014), \texttt{1403.4788}

\bibitem{Christiansen:2015jpa}
J.R. Christiansen, S.~Prestel (2015), \texttt{1510.01517}

\bibitem{Gieseke:2014gka}
S.~Gieseke, T.~Kasprzik, J.H. K{\"u}hn, Eur. Phys. J. \textbf{C74}, 2988
  (2014), \texttt{1401.3964}

\bibitem{Kallweit:2014xda}
S.~Kallweit, J.M. Lindert, P.~Maierh{\"o}fer, S.~Pozzorini, M.~Sch{\"o}nherr,
  JHEP \textbf{04}, 012 (2015), \texttt{1412.5157}

\bibitem{Frixione:2015zaa}
S.~Frixione, V.~Hirschi, D.~Pagani, H.S. Shao, M.~Zaro, JHEP \textbf{06}, 184
  (2015), \texttt{1504.03446}

\bibitem{Kallweit:2015dum}
S.~Kallweit, J.M. Lindert, S.~Pozzorini, M.~Sch{\"o}nherr, P.~Maierh{\"o}fer
  (2015), \texttt{1511.08692}

\bibitem{Frixione:2002ik}
S.~Frixione, B.R. Webber, JHEP \textbf{06}, 029 (2002), \texttt{hep-ph/0204244}

\bibitem{Nason:2004rx}
P.~Nason, JHEP \textbf{11}, 040 (2004), \texttt{hep-ph/0409146}

\bibitem{Frixione:2007nw}
S.~Frixione, P.~Nason, G.~Ridolfi, JHEP \textbf{09}, 126 (2007),
  \texttt{0707.3088}

\bibitem{Hoeche:2012ft}
S.~H{\"o}che, F.~Krauss, M.~Sch{\"o}nherr, F.~Siegert, Phys. Rev. Lett.
  \textbf{110}, 052001 (2013), \texttt{1201.5882}

\bibitem{Hoche:2012wh}
S.~H{\"o}che, M.~Sch{\"o}nherr, Phys. Rev. \textbf{D86}, 094042 (2012),
  \texttt{1208.2815}

\bibitem{Cascioli:2013era}
F.~Cascioli, P.~Maierh{\"o}fer, N.~Moretti, S.~Pozzorini, F.~Siegert, Phys.
  Lett. \textbf{B734}, 210 (2014), \texttt{1309.5912}

\bibitem{Hoche:2010pf}
S.~H{\"o}che, F.~Krauss, M.~Sch{\"o}nherr, F.~Siegert, JHEP \textbf{04}, 024
  (2011), \texttt{1008.5399}

\bibitem{Czakon:2015cla}
M.~Czakon, H.B. Hartanto, M.~Kraus, M.~Worek, JHEP \textbf{06}, 033 (2015),
  \texttt{1502.00925}

\bibitem{Lonnblad:2012ix}
L.~L{\"o}nnblad, S.~Prestel, JHEP \textbf{03}, 166 (2013), \texttt{1211.7278}

\bibitem{Hamilton:2013fea}
K.~Hamilton, P.~Nason, E.~Re, G.~Zanderighi, JHEP \textbf{10}, 222 (2013),
  \texttt{1309.0017}

\bibitem{Hoche:2014dla}
S.~H{\"o}che, Y.~Li, S.~Prestel, Phys. Rev. \textbf{D90}, 054011 (2014),
  \texttt{1407.3773}

\bibitem{Catani:2001cc}
S.~Catani, F.~Krauss, R.~Kuhn, B.R. Webber, JHEP \textbf{11}, 063 (2001),
  \texttt{hep-ph/0109231}

\bibitem{Hamilton:2012np}
K.~Hamilton, P.~Nason, G.~Zanderighi, JHEP \textbf{10}, 155 (2012),
  \texttt{1206.3572}

\bibitem{Hamilton:2012rf}
K.~Hamilton, P.~Nason, C.~Oleari, G.~Zanderighi, JHEP \textbf{05}, 082 (2013),
  \texttt{1212.4504}

\bibitem{Karlberg:2014qua}
A.~Karlberg, E.~Re, G.~Zanderighi, JHEP \textbf{09}, 134 (2014),
  \texttt{1407.2940}

\bibitem{Hoeche:2014aia}
S.~H{\"o}che, Y.~Li, S.~Prestel, Phys. Rev. \textbf{D91}, 074015 (2015),
  \texttt{1405.3607}

\bibitem{Frederix:2012ps}
R.~Frederix, S.~Frixione, JHEP \textbf{12}, 061 (2012), \texttt{1209.6215}

\bibitem{Hoeche:2014qda}
S.~H{\"o}che, F.~Krauss, P.~Maierh{\"o}fer, S.~Pozzorini, M.~Sch{\"o}nherr,
  F.~Siegert, Phys. Lett. \textbf{B748}, 74 (2015), \texttt{1402.6293}

\bibitem{Lavesson:2008ah}
N.~Lavesson, L.~Lonnblad, JHEP \textbf{12}, 070 (2008), \texttt{0811.2912}

\bibitem{Hoeche:2012yf}
S.~H{\"o}che, F.~Krauss, M.~Sch{\"o}nherr, F.~Siegert, JHEP \textbf{04}, 027
  (2013), \texttt{1207.5030}

\bibitem{Gehrmann:2012yg}
T.~Gehrmann, S.~H{\"o}che, F.~Krauss, M.~Sch{\"o}nherr, F.~Siegert, JHEP
  \textbf{01}, 144 (2013), \texttt{1207.5031}

\bibitem{Hoche:2010kg}
S.~H{\"o}che, F.~Krauss, M.~Sch{\"o}nherr, F.~Siegert, JHEP \textbf{08}, 123
  (2011), \texttt{1009.1127}

\bibitem{Cascioli:2013gfa}
F.~Cascioli, S.~H{\"o}che, F.~Krauss, P.~Maierh{\"o}fer, S.~Pozzorini,
  F.~Siegert, JHEP \textbf{01}, 046 (2014), \texttt{1309.0500}

\bibitem{Hoeche:2014lxa}
S.~H{\"o}che, F.~Krauss, M.~Sch{\"o}nherr, Phys. Rev. \textbf{D90}, 014012
  (2014), \texttt{1401.7971}

\bibitem{Hoeche:2014rya}
S.~H{\"o}che, F.~Krauss, S.~Pozzorini, M.~Sch{\"o}nherr, J.M. Thompson, K.C.
  Zapp, Phys. Rev. \textbf{D89}, 093015 (2014), \texttt{1403.7516}

\bibitem{Buschmann:2014sia}
M.~Buschmann, D.~Goncalves, S.~Kuttimalai, M.~Sch{\"o}nherr, F.~Krauss,
  T.~Plehn, JHEP \textbf{02}, 038 (2015), \texttt{1410.5806}

\end{thebibliography}

\end{document}